\documentclass{desyprocA4}
\newcommand\pubdate{\today}

\def\to{\rightarrow}

\def\bi{\begin{itemize}}
 \def\ei{\end{itemize}}

\def\c1p{C1^\prime}
\def\msq3{\overline{m}_{\tilde{q}}(3)}

\def\tst{\tilde t}

\def\be{\begin{equation}}  
\def\ee{\end{equation}}  
\def\bea{\begin{eqnarray}}  
\def\eea{\end{eqnarray}}  
\def\tw{\widetilde W}

\def\tz{\tilde Z}

\def\beq{\begin{equation}}
\def\eeq#1{\label{#1}\end{equation}}
\def\eeqn{\end{equation}}

\newenvironment{Eqnarray}%
   {\arraycolsep 0.14em\begin{eqnarray}}{\end{eqnarray}}
\def\beqa{\begin{Eqnarray}}
\def\eeqa#1{\label{#1}\end{Eqnarray}}
\def\eeqan{\end{Eqnarray}}

\begin{document}
\title{Naturalness, Supersymmetry and Light Higgsinos:\\
A Snowmass Whitepaper}

\author{{\slshape Howard Baer$^1$, Vernon Barger$^2$, Peisi Huang$^2$, Dan
    Mickelson$^1$, Azar Mustafayev$^3$ and Xerxes Tata$^3$},\\
  $^1$Dept. of Physics and Astronomy, University of Oklahoma, Norman, OK 73019, USA\\ 
$^2$ Dept. of Physics, University of Wisconsin, Madison, WI 53706, USA\\
$^3$Dept. of Physics and Astronomy, University of Hawaii at Manoa,
    Honolulu, HI 96822, USA}


\maketitle

\pubdate

\begin{abstract}
We show that the electroweak fine-tuning parameter $\Delta_{\rm EW}$
derived from the well-known electroweak symmetry breaking condition written
in terms of weak scale parameters leads to {\it a bound on fine-tuning
in the MSSM} and explain its utility for phenomenological analyses. We
argue that a small magnitude of the $\mu$ parameter, and the concomitant
presence of light higgsinos, is the most basic consequence of naturalness
in SUSY models, and list the resulting implications of this for experiments at
the LHC and at future $e^+e^-$ colliders.
\end{abstract}

%

The ultra-violet behaviour of softly broken SUSY theories, with SUSY
broken at the weak scale, ensures that the low energy theory is at most
logarithmically sensitive to high scale (HS) physics. Thus weak scale
SUSY solves the {\it big hierarchy problem} endemic to the Standard
Model (SM) and discussions of naturalness concern at most the
logarithmic sensitivity to HS physics.

The well-known electroweak symmetry breaking condition that yields the
value of $M_Z$ using the one-loop renormalization group improved
effective potential can be written as,
\be \frac{M_Z^2}{2} =
\frac{m_{H_d}^2 + \Sigma_d^d -
(m_{H_u}^2+\Sigma_u^u)\tan^2\beta}{\tan^2\beta -1} -\mu^2 \;.
\label{eq:loopmin}
\ee 
Here, $\Sigma_u^u$ and $\Sigma_d^d$ are the one-loop corrections
arising from loops of particles and their superpartners that couple
directly to the Higgs doublets. Naturalness requires that there be no large 
cancellations between terms on the right-hand-side of (\ref{eq:loopmin}). We are thus
led to require that \cite{rnslet}
\be 
 \Delta_{\rm EW} =
max\left( \left|\frac{-m_{H_u}^2\tan^2\beta}{\tan^2\beta-1}\right|,\left|\frac{-\Sigma_u^u (\tst_1 )\tan^2\beta}{\tan^2\beta-1}\right|,
\cdots,|-\mu^2| \right)/(M_Z^2/2)
\label{eq:ewft}
\ee 
remain smaller than some pre-chosen value decided by how much fine-tuning one ``deems reasonable''.

We note that (\ref{eq:ewft}) is obtained from the {\it weak scale} Lagrangian
and so contains no information about either the underlying HS theory or
the logarithms that we mentioned above. To see these logs, we must rewrite
$m_{H_u}^2$, $m_{H_d}^2$ and $\mu^2$ in (\ref{eq:loopmin}) in terms of
these parameters defined at the high scale using $m_{H_u}^2 =
m_{H_u}^2(\Lambda) + \delta m_{H_u}^2$, {\it etc.}: the large logarithms
are then contained in $\delta m_{H_u, H_d}^2$ and in $\delta\mu^2$. We
can then define a fine-tuning measure $\Delta_{\rm HS}$, that ``knows
about the high energy theory'' in an analogous manner to $\Delta_{\rm
EW}$ \cite{sugra}. Typically, $\delta m_{H_u}^2$ has dominant
contributions from top squark loops; this has led many authors to argue
that top squarks must be lighter than 400-600~GeV in natural SUSY models
with $\Lambda$ as low as $\sim 10$ TeV\cite{kn,pap} (limits are even lower
for higher $\Lambda$).  Note that $\Delta_{\rm HS}$ is a very strict measure of
naturalness in that any cancellations between $m_{H_u}^2(\Lambda)$ and
$\delta m_{H_u}^2$ that lead to low $m_{H_u}^2$ at the weak scale will
lead to a large value of $\Delta_{\rm HS}$, but not of $\Delta_{\rm
EW}$.

The reader may correctly note that $\Delta_{\rm HS}$ as we have
defined it does not incorporate correlations amongst weak scale
parameters that are inevitably present in HS models such as mSUGRA/CMSSM that
are completely specified by a small number of HS parameters. Because of
these correlations, the weak scale value of $m_{H_u}^2$ can be
approximated (using the one-loop solutions of the RGEs for
$\tan\beta=10$ and working within mSUGRA/CMSSM) by \cite{fp},
\be
-2m_{H_u}^2(m_{\rm weak}) = 3.78m_{1/2}^2-0.82 A_0 m_{1/2} + 0.22A_0^2
+0.013m_0^2\;,
\label{eq:corr}
\ee
with analogous expressions for $m_{H_d}^2$ and $\mu^2$ that can be
substituted in (\ref{eq:loopmin}).  The small coefficient of $m_0^2$,
which arises mainly because of the underlying equality of the HS soft SUSY
breaking mass parameters for the $H_u$, $Q_3$ and $U_3$ fields (see
Eq.~(28) of Ref. \cite{fp}), then implies that the fine-tuning in
mSUGRA/CMSSM may be smaller than expected for large values of $m_0$, as
long as $m_{1/2}$ is not very large. We remark, however, that in this
region -- the focus point/hyperbolic branch region of mSUGRA/CMSSM --
$\Delta_{\rm EW}$ becomes large for very large $m_0$ because the
$\Sigma_u^u$ in (\ref{eq:ewft}) begins to dominate.  Note that
incorporating the correlations between parameters as in (\ref{eq:corr})
allows a (partial) cancellation between $m_{H_u}^2(\Lambda)$ and $\delta
m_{H_u}^2$ {\it without a concomitant contribution to the fine-tuning
measure}, so that this criterion is intermediate between $\Delta_{\rm
HS}$ and $\Delta_{\rm EW}$. Because of the simple quadratic form in the
expression for $M_Z^2$ that results in this approximation, these
considerations lead to a fine-tuning measure numerically similar to the
widely used Barbieri-Guidice measure $\Delta_{\rm
BG}\equiv\frac{\partial\ln M_Z^2}{\partial\ln a_i}$ \cite{EENZ,BG} where
the $a_i$ are fundamental theory parameters.  We thus have,
$$\Delta_{\rm EW}< \Delta_{\rm BG}{\lesssim} \Delta_{\rm HS}.$$

While $\Delta_{\rm HS}$ is a popular (albeit strict) fine-tuning measure
in a HS theory and $\Delta_{BG}$ is a traditional one, it is reasonable
to wonder about the utility of $\Delta_{\rm EW}$, a quantity that is
independent of the HS physics. We find that: $\Delta_{\rm EW}$ is
clearly {\it a bound on the fine-tuning}\cite{rns}.
\footnote{ In HS effective theories that  are obtained  from an overarching 
meta-theory that fixes the parameters of the HS theory so that
cancellations between $m_{H_u}^2(\Lambda)$ and the log terms in $\delta
m_{H_u}^2$ are {\it automatic}, we would not pay a fine-tuning price for
the cancellation and
$\Delta_{\rm EW}$ would be a valid fine-tuning measure.}  We can use it
to infer \cite{sugra} that models with large $\Delta_{\rm EW}$ (such as
mSUGRA/CMSSM where $\Delta_{EW}$ always exceeds $\sim 100$) 
are necessarily fine-tuned.\footnote{Implicit in this is an
assumption that the SUSY conserving $\mu$ parameter and the soft SUSY
breaking parameters have different origin, and so cancellations between
these are unnatural.}  Moreover, the properties of $\Delta_{\rm EW}$
allow us to make important phenomenological inferences.
\bi
\item $\Delta_{\rm EW}$ is essentially determined by the physical
  spectrum. Thus, in principle, if a low energy spectrum leads to a
  large value of $\Delta_{\rm EW}$ we could infer the underlying theory
  is necessarily fine-tuned. {\it A low value of $\Delta_{\rm EW}$ does
  not imply the absence of fine-tuning,} but leaves open the possibility
  that there is an underlying theory with this spectrum that may not be
  fine-tuned. Since many observable cosequences are determined
  largely by the spectrum, we can study the phenomenology of these
  theories even without detailed knowledge of the underlying HS theory!

\item Low $\Delta_{\rm EW}< 10\ (30)$ necessarily implies small $|\mu|<
  200\ (300)$ GeV, and hence the existence of light higgsino states.  An
  $e^+e^-$ collider operating with $\sqrt{s}\gtrsim 2|\mu |$ may be
  able to see the low visible-energy-release events from
  $\tw_1^+\tw_1^-$, $\tz_1\tz_2$ and $\tz_2\tz_2$ higgsino pair production 
  above SM   two-photon backgrounds (as indicated by early analyses
  \cite{krupov}). Thus, $e^+e^-$ colliders will decisively probe
  $\Delta_{\rm EW}<\frac{1}{2}\left(\frac{E_{\rm CM}}{M_Z}\right)^2$.

\item It is possible to find models with low values of $\Delta_{\rm EW} \sim 10-30$
  but with top-squarks in the 1-5~TeV range \cite{rns}. For this reason, {\it
  we regard small $|\mu|$ as a more basic consequence of naturalness considerations
  than sub-TeV top squarks}. The importance of small $|\mu|$, though
  present, remains obscured in  analyses that focus on the large
  top squark contributions to $\delta m_{H_u}^2$ mentioned above \cite{pap}.

\item If $|M_{1,2}|$ is coincidently comparable to $|\mu|$, charginos
and neutralinos would be highly mixed gaugino-higgsino states and their
electroweak production would lead to observable multilepton production
at the LHC.

\item If $|M_2| \le 0.8-1$~TeV, but $|\mu|$ is small, hadronically quiet
  same-sign dilepton events from wino pair production $\tw_2\tz_4 \to
  W^\pm W^\pm+E_T^{\rm miss}$ may offer the greatest SUSY reach at a
  high luminosity LHC \cite{newlhc}.

\item If gaugino masses are large, the small mass gap between
  $\tw_1-\tz_2$ and $\tz_2 -\tz_1$ ($\sim 10-15$ GeV) makes detection of
  higgsino pair production exceedingly difficult to see at the LHC
  because the visible decay products are very soft. In this case, it
  would be worth examining the initial state monojet and mono-photon
  radiation signal from higgsino pair production at LHC to see if they
  are possible avenues for discovery.  \ei

\section*{Acknowledgments} 
This research was sponsored in part by grants from the US Department of Energy


\begin{footnotesize}

\end{footnotesize}


\end{document}